\documentclass[10pt]{article}

\usepackage{amsmath}
\usepackage{amssymb}
\usepackage{graphics}
\usepackage{rotating}
\usepackage{cite}
\usepackage{color}
\usepackage{fancybox}

\def\0{\mbox{\tiny $0$}}
\def\1{\mbox{\tiny $1$}}
\def\2{\mbox{\tiny $2$}}
\def\3{\mbox{\tiny $3$}}
\def\4{\mbox{\tiny $4$}}
\def\5{\mbox{\tiny $5$}}
\def\6{\mbox{\tiny $6$}}
\def\7{\mbox{\tiny $7$}}
\def\8{\mbox{\tiny $8$}}
\def\9{\mbox{\tiny $9$}}

\def\NR{\mbox{\tiny $NR$}}
\def\I{\mbox{\tiny $I$}}

\title{\shadowbox{\large \bf ANTIPARTICLES CREATION IN TUNNELLING }}

\author{
\small  Stefano De Leo\thanks{Department of Applied Mathematics,
State University of Campinas, Brazil [deleo@ime.unicamp.br] } \,\,
and\, Pietro Rotelli\thanks{Department of Physics, University of
Salento and INFN Lecce, Italy [rotelli@le.infn.it]}}

\date{\small
\fcolorbox{black}{yellow} {\color{red} $\bullet$ {\color{black}{
{\footnotesize  {\sc International Journal of Modern Physics A} {\bf 28}, 1350129-9 (2013)}}
{\color{red}{$\bullet$}} } }}

\begin{document}
%

\maketitle

\vspace*{-.7cm}


\begin{abstract}
\noindent We study particle interaction with a Dirac step
potential. In the standard Klein energy zone, the hypothesis of
Klein pair production predicts the existence of free/oscillatory
antiparticles. In this paper, we discuss the tunneling energy zone
characterized by {\em evanescent} wave functions in the
classically forbidden region. We ask the question of the nature,
particle or antiparticle, of the densities within the classically
forbidden region. The answer to this question is relevant to the
correct form of the reflection coefficient. \end{abstract}











\section*{\normalsize I. INTRODUCTION}
 When considering the Dirac equation for a step potential,
\[
V(z) = \left\{
\begin{array}{cll} 0 & \,\,\,\,\,\mbox{for}\,\, z<0&\,\,\,\,\, (\mbox{region
I})\,\,,\\
V_{\0} & \,\,\,\,\,\mbox{for}\,\,z>0&\,\,\,\,\, (\mbox{region
II})\,\,,
\end{array}
\right.
\]
three distinct energy zones are evident\cite{ZUB,GRO},
\begin{center}
\begin{tabular}{lcl}
zone 1 & - & $E>V_{\0}+m$\,\,\,\,\,(diffusion) \,\,,\\
zone 2 & -  & $V_{\0}-m<E<V_{\0}+m$\,\,\,\,\,(tunneling)\,\,,\\
zone 3 & - & $E<V_{\0}-m$\,\,\,\,\,(Klein pair
production\cite{KLE})\,\,.
\end{tabular}
\end{center}
We are using here standard barrier language (tunneling) extended
to the step potential. The argument for pair
production\cite{KLE1,KLE2,KLE3,DELK} will be re-derived below. In
a previous paper\cite{DELK}, we studied the Klein energy zone 3 in
which it has been hypothesized that pair production occurs. The
reflected beam is of the same nature (particle) as that of the
incoming beam, while the created antiparticles see a well
potential and consequently travel forward freely. Wave packets can
be formed and group velocities defined. In this paper, we consider
the tunneling energy zone characterized by evanescent (non free)
solutions in the classical forbidden region. For these solutions
currents (and consequently group velocities) do not exist. One of
the conclusions of this work will be that the tunneling energy
zone must be considered as two separate tunneling zones,
\begin{center}
\begin{tabular}{lcl}
zone 2a & - & $V_{\0}<E<V_{\0}+m$\,\,,\\
zone 2b & - &  $V_{\0}-m<E<V_{\0}$\,\,.
\end{tabular}
\end{center}
The distinction will be the nature of the ``particles" within the
step. We shall argue that in zone 2a they are particles, while
within 2b they are antiparticles. This will necessarily require a
modification of the reflection coefficient $R$ which will be
discontinuous in phase at $E=V_{\0}$.

Since the Dirac equation is a spinor equation\cite{ZUB}, let us
simplify our language by referring to the particles as electrons
and the antiparticles as positrons. We assume incoming electrons
from the left. As is standard, we work analytically with plane
waves, but at some point we will perform numerical calculations
using gaussian wave packets.

In the next section, we recall the results for diffusion.
Subsequently, in section III, we pass to what we wish to call the
Dirac tunneling (zone 2a). We observe there the analytical
connection between the diffusion and tunneling results. IN section
IV, we jump to the Klein energy zone where pair production is
assumed and derive the reflection and transmission amplitudes. In
analogy with the diffusion-Dirac tunneling relationship, we derive
the Klein tunneling results in section IV. Our conclusions, in
particular the nature of the ``particles'' in the classically
forbidden region, are drawn in the final section.

Throughout the paper extensive use is made of charge density in
the free region as a function of time. The increase or decrease
during reflection (transition period) is an excellent indicator of
the charge of particles under the step.

\section*{\normalsize II. DIFFUSION $\boldsymbol{E>V_{\0}+m}$}

Let the incoming electrons be spin up, this choice does not
influence our results. Continuity at $z=0$ reads
\begin{equation}
\label{con} u(p,E) + R\,u(-p,E) = T\,u(q,E-V_{\0}) \,\,,
\end{equation}
where $u(p,E) = \left[\,1\,,\,0\,,\,p/(E+m) \,,\,0  \,\right]^t$,
$p=\sqrt{E^{^{2}}-m^{\2}}$ and\,
$q=\sqrt{(E-V_{\0})^{^{2}}-m^{\2}}$. We have, for simplicity,
absorbed the spinor normalizations ratio within $T$. Solving for
$R$ and $T$, we find
\begin{equation}
R=\frac{1-\alpha}{1+\alpha}\,\,\,\,\,\mbox{and}\,\,\,\,\,T=\frac{2}{1+\alpha}\,\,,
\end{equation}
where  $\alpha=q\,(E+m)\,/\,p\,(E-V_{\0}+m)\,>\,0$. There is no
room for spin flip\cite{DELD1,DELD2}. Obviously the fermions in
region II are electrons  because flux conservation requires this.
Indeed flux conservation with our choices for $R$ and $T$ implies
\begin{equation}
|R|^{^{2}}+\alpha\,|T|^{^{2}}=1\,\,.
\end{equation}
Observe that of course $R<1$, with $R\to 0$ for $E \to \infty$,
but $R\to 1$ for $E \to V_{\0}+m$. In this limit ($\alpha\to 0$)
there is no flux in region II, but that is because the electron
velocity has gone to zero. We point out that there is nevertheless
an electron density  proportional to  $|T|^{^{2}}(\to 4$ in this
limit).

\section*{\normalsize III. DIRAC TUNNELING $\boldsymbol{V_{\0}<E<V_{\0}+m}$}

The spatial dependence in region II is now
$\exp[-\,\widetilde{q}\,z]$ with
$\widetilde{q}=\sqrt{m^{\2}-(E-V_{\0})^{^{2}}}>0$. Thus,
$\widetilde{q}^{\,\2}=-\,q^{\2}$. For diffusion it was of course
$\exp[i\,q\,z]$. This readily suggests how we must modify our
spinors in the continuity equations. Formally $q^{\2}$ is now
negative whence $q$ is now imaginary. To pass from the oscillatory
behavior in diffusion to the evanescent behavior, we simply
perform $q\to i\, \widetilde{q}$. This must be done also within
the spinor. Solving the continuity equations now yields
\begin{equation}
R=\frac{1-i\,\widetilde{\alpha}}{1+i\,\widetilde{\alpha}}\,\,\,\,\,\mbox{and}\,\,\,\,\,
T=\frac{2}{1+i\,\widetilde{\alpha}}\,\,,
\end{equation}
where
$\widetilde{\alpha}=\widetilde{q}\,(E+m)\,/\,p\,(E-V_{\0}-m)\,>\,0$.

The well known feature of this result is that $|R|=1$, total
reflection occurs consistent with the fact that while there are
fermions within the step there is no flux since the spatial
dependence is evanescent, more specifically exponentially
decreasing with increasing $z$.

Now, we are tempted both by continuity of particle density in the
transition from diffusion to/from tunneling to identify these
fermions as electrons. However, there exists a consistency check
involving wave packets in region I. With wave packets (we shall
use gaussian momentum distributions) the solution of the Dirac
equation is no longer stationary but represents an incident right
moving wave packet for $t\ll 0$ and a reflected left moving wave
packet for $t\gg 0$. This situation could have been ``read" from
our plane wave results. The important difference between plane
waves vs. wave packets occurs when considering $t\approx 0$. This
is a period of transition for wave packets during which fermion
density in region II (under the step) first grows from zero ($t\ll
0$) and then returns to zero ($t\gg 0$). During this transition it
is legitimate to ask how the total numbers of electrons in region
I changes. We have discovered that this depends crucially upon the
analytic expression for the reflection coefficient.

In Fig.\,1b, we show the ratio of this density to the incoming
density as a function of time. That is, we have plotted
\begin{equation}
\label{rt} r(t)=\int_{\mbox{\tiny $-
\infty$}}^{\,\0}\hspace*{-.15cm} \mbox{d}z\,\left|\Psi_{\I}(z,t)
\right|^{^{2}}\,\mbox{\huge$/$}\, \int_{\mbox{\tiny $-
\infty$}}^{\,\0}\hspace*{-.15cm}
\mbox{d}z\,\left|\Psi_{\I}(z,-\infty) \right|^{^{2}}\,\,,
\end{equation}
where
\[
\Psi_{\I}(z,t) = \int_{\0}^{w}\hspace*{-.15cm} \mbox{d}p\,\,g(p)\,
\left[\,u(p,E)\,e^{ipz} + R\,u(-p,E)\,e^{-ipz}\,
\right]\,e^{-iEt}\,\,,
\]
with $g(p)=\exp[(p-p_{\0})^{^{2}}d^{^{2}}/4]$ and $w$ is the
maximum value of $p$ compatible with the tunneling energy zone,
i.e. $w=\sqrt{V_{\0}(V_{\0}+2m)}$, and $p_{\0}<w$ the chosen value
of peak incoming momentum. As can be seen from the plot, the value
of $r(t)$ is at all times $\leq 1$. During transition there is a
loss of electrons in free space (region I) which necessarily
implies that the fermions within region II must be dominantly
electrons\cite{DELT}. Invoking total normalization conservation
implies they are all electrons. Charge conservation is hence
conserved as it must be.

We would considered the non relativistic limit, $m\gg
V_{\0}>E_{\NR}=E-m$, and consequently the Schr\"odinger equation
approximation. With Schr\"odinger  antiparticle production is
ignored. Consequently, we would have automatically assumed thet
the objects within the classically forbidden region were
electrons. However, this would not have been by itself a  proof.
Before considering the other tunneling energy zone, we jump to the
Klein zone and its interpretation.

\section*{\normalsize IV. KLEIN PAIR PRODUCTION $\boldsymbol{E<V_{\0}-m}$}

This energy zone is characterized by oscillatory solutions in
region II as occurs for diffusion\cite{KLE}. Indeed,
$E-V_{\0}<-m$, and even if $E-V_{\0}$ is negative, this also
implies a real $q$ as for diffusion. Again the continuity equation
is given by Eq.\,(\ref{con}) if the solution in region II is
chosen to be the plane wave $u(q,E-V_{\0})\,\exp[i(qz-Et)]$. The
essential difference here compared to diffusion is that $\alpha$
is now negative, consequently $|R|>1$. This fact implies that more
electrons are reflected than those incident, see Fig.\,1d.

Mathematically, this is in accord with the observation that the
wave in region II has a negative group velocity, $q/(E-V_{\0})$,
i.e. it represents ``electrons'' incident  from the right. Where
one to assume the alternative oscillatory behavior $\exp[-iqz]$
one would have obtained  $|R|<1$ as in diffusion. However, Klein
observed that the only physical interpretation for ``free"
fermions in region II is that they be positrons. Since the
potential is subtracted from the energy, it is an electrostatic
potential. Consequently, {\em if an ``electron" sees a potential
$V_{\0}$, a positron sees a potential $-V_{\0}$}. Its wave
function is the complex conjugate of the ``electron'' wave
function (charge conjugation), so the {\em physical} positron wave
function will be proportional to $\exp[- i(q_az-E_at)]$ where
$q_a=\sqrt{[E_a-(-V_{\0})]^{^{2}}-m^{\2}}$ with $E_a$ the positron
energy. To have a common time dependence, $\exp[-iEt]$
(essentially for the continuity equations), we must have $E_a=-E$,
which while negative is nevertheless above the potential
($-V_{\0}$) seen by the positron. Indeed $-E>-V_{\0}+m$ so it
represents free positrons. The group velocity,
$q_a/[E_a-(-V_{\0})] =q/(V_{\0}-E)$, is now positive and,
consequently, the flux of positrons will be from left to right and
charge conservation will hold although conserved fermion density
will not. The excess of reflected electrons is equal to the
positrons created at the potential discontinuity. We have pair
production. This interpretation involving Klein pair production is
quite conventional\cite{KLE1,KLE2,KLE3}. The alternative with
$|R|<1$ in the Klein zone would correspond to incoming
antiparticle flux from the right and pair annihilation with some
of the incident electrons. In either case, positrons are needed to
explain the oscillatory behavior in region II.

We wish to recall a few facts about this particular form of pair
production. Since the electrons/positrons live in separate regions
with diverse potentials, the creation process is achieved with
zero net energy cost, zero net current, and zero net
helicity\cite{DELK}.

\section*{\normalsize V. KLEIN TUNNELING $\boldsymbol{V_{\0}-m<E<V_{\0}}$}

Now, to analytically continue from the Klein zone into what we
have labelled the Klein tunneling zone 2b, we make the hypothesis
that the spatial wave function to consider in the Klein zone,
$E<V_{\0}-m$, is $\exp[- iq_az]=\exp[- iq z]$. As we argued above
when passing from the diffusion zone to the tunneling zone, which
we called the Dirac tunneling zone, we must have in region II the
form $\exp[-\widetilde{q}z]$. This is now achieved by
$q\to-i\widetilde{q}$. Whence
\begin{equation}
R=\frac{1+i\,\widetilde{\alpha}}{1-i\,\widetilde{\alpha}}\,\,\,\,\,\mbox{and}\,\,\,\,\,
T=\frac{2}{1-i\,\widetilde{\alpha}}\,\,,
\end{equation}
with $\widetilde{\alpha}$ as defined above. Formally, this $R$ is
the complex conjugate of the previous Dirac tunneling expression.
However, we must always remember that the two expressions are
valid in different energy zones (2a and 2b). Again $|R|=1$
implying total reflection also in this tunneling zone.

Now, we perform our wave packet analysis. In Fig.\,1c, we see that
during transition there is an excess of electrons in region I.
This implies, with the analogous argument to that for the Dirac
tunneling zone, that the fermions in region II are positrons. This
is in accordance with the feature that as we pass from Klein to
tunneling, at the energy interface $E=V_{\0}-m$, there will be a
constant density of stationary positrons in region II. Here
continuity is maintained again in perfect analogy with what
happens for electrons when passing from diffusion to Dirac
tunneling. Had we chosen to follow the alternative route, $q\to
i\widetilde{q}$, we would have predicted below potential electrons
in region II. In this case the reflection coefficient would have
been the complex conjugate to that given above, and formally
coincided with the $R$ for Dirac tunneling.

In Fig.\,2, we plot the values of Arg[$R$] and $|R|$ vs. $E/m$ for
the case $V_{\0}=3.5\,m$. The tunneling zone is clearly identified
by $|R|=1$ and consists of what we have labelled Klein tunneling
(KT) and Dirac tunneling (DT). Note the discontinuity in phase at
$E=V_{\0}$. We also observe that there is a peak value to Klein
pair production ($|R|>1$) at $E=V_{\0}/2$.

\section*{\normalsize VI. CONCLUSIONS}

For our step analysis, we have divided the incoming particle range
$E$ into four sub-ranges separated at the values $V_{\0}+m$,
$V_{\0}$, and $V_{\0}-m$ (obviously the last interface requires
 $V_{\0}>2\,m$ since always
$E\geq  m$). In the higher band, $E>V_{\0}+m$, we have diffusion
characterized by a reflection coefficient $|R|<1$. In the second,
we encounter a tunneling zone although for a step (or in practice
for a very large barrier) no tunneling actually occurs.  In this
energy zone $|R|=1$, but nevertheless electrons exist within the
classically forbidden region II. We then jumped to the Klein zone
$E<V_{\0}-m$ and recalled the interpretation of the ``free"
particles (oscillatory spatial wave function) with positrons
travelling over a potential of $-V_{\0}$. Extrapolating $E$
upwards ($E_a$ downwards) into the 2b tunneling zone, we derived a
new $R$ still with $|R|=1$ but consistent with  positrons in
region II.

Now, we wish to make two technical points.

1) In this paper, we have used exclusively the spinors
$u^{\mbox{\tiny $(1)$}}$. However, for $E<V_{\0}$, we have
$E-V_{\0}<0$. Comparing with the study of free fermions when
$V_{\0}=0$, we are in the realm of ``negative energy". We have
argued elsewhere\cite{DELK} that it is thus more logical to use
$u^{\mbox{\tiny $(3)$}}$ for $E<V_{\0}$. The fact is that,
surprisingly, this would change none of our results. So, we stick
to $u^{\mbox{\tiny $(1)$}}$ for simplicity, in accordance with
most literature including Klein himself\cite{KLE,KLE1,KLE2,KLE3}.
Actually, within the Klein zone, we have above potential positrons
so there the use of $u^{\mbox{\tiny $(1)$}}$ is fully justified.

2) The creation of positrons  implies that the two step approach
to the analysis of say a barrier (or multiple step for a general
potential structure) is not the same as the standard solution
involving coupled matrix equations. For example, the positrons
created {\em alla} Klein in a barrier (potential well for them)
are permanently trapped therein\cite{DELK}. This feature would not
be evident in the standard procedure. Indeed, the standard
solution for a barrier in the Klein energy zone gives $|R|<1$.
This fact has generated some doubts, even recently, with Klein's
interpretation. We are of the opinion that when the solutions
differ the multiple step solution is the only consistent one.
Other solutions can be traced to the summation of non convergent
series.

The main conclusion of this paper is that the tunneling zone must
be divided into two parts, 2a ($V_{\0}<E<V_{ \0}+m$) and 2b
($V_{\0}-m<E<V_{ \0}$). In the former the fermions in region II
are electrons, in the latter they are positrons. Pair creation is
thus predicted even for below potential conditions. This extends
the Klein pair production to this part of the tunneling energy
zone. The appropriate $R$ function in zone 2b has been found and
it is formally the complex conjugate of that in zone 2a. There is
a consequent discontinuity at $E=V_{\0}$ both for the phase of $R$
and for the nature of the fermions. Another byproduct of this
phase change is that in DT the reflected wave packets are time
delayed while in KT they are time advanced. Obviously, these
conclusions are subject to eventual experimental verification,
possibly by the measurement of the ratio of the reflected
particles to the incoming particles in region I, defined by the
observable $r(t)$ given in Eq.(\ref{rt}). To the best of our
knowledge even the creation of Klein pairs has yet to be verified
experimentally. Pair creation is of course basic to field
theory\cite{GRO}, and that is why Klein's hypothesis is considered
a precursor to field theory.

Recent developments in graphene physics\cite{NOS1,NOS2,NOS3} have
shown that a Dirac like excitation with a zero mass occurs. This
opens up a very practice possibility of testing both Klein pair
productions and tunneling aniparticles proposed in this paper.

\section*{\small \rm ACKNOWLEDGEMENTS}

One of the authors (SdL) thanks the Department of Physics
(University of Salento, Italy) for the hospitality and the FAPESP
(Brazil) for financial support by the Grant No. 10/02213-3.

\newpage

\begin{figure}[hbp]
\hspace*{-2.5cm}
\includegraphics[width=19cm, height=24cm, angle=0]{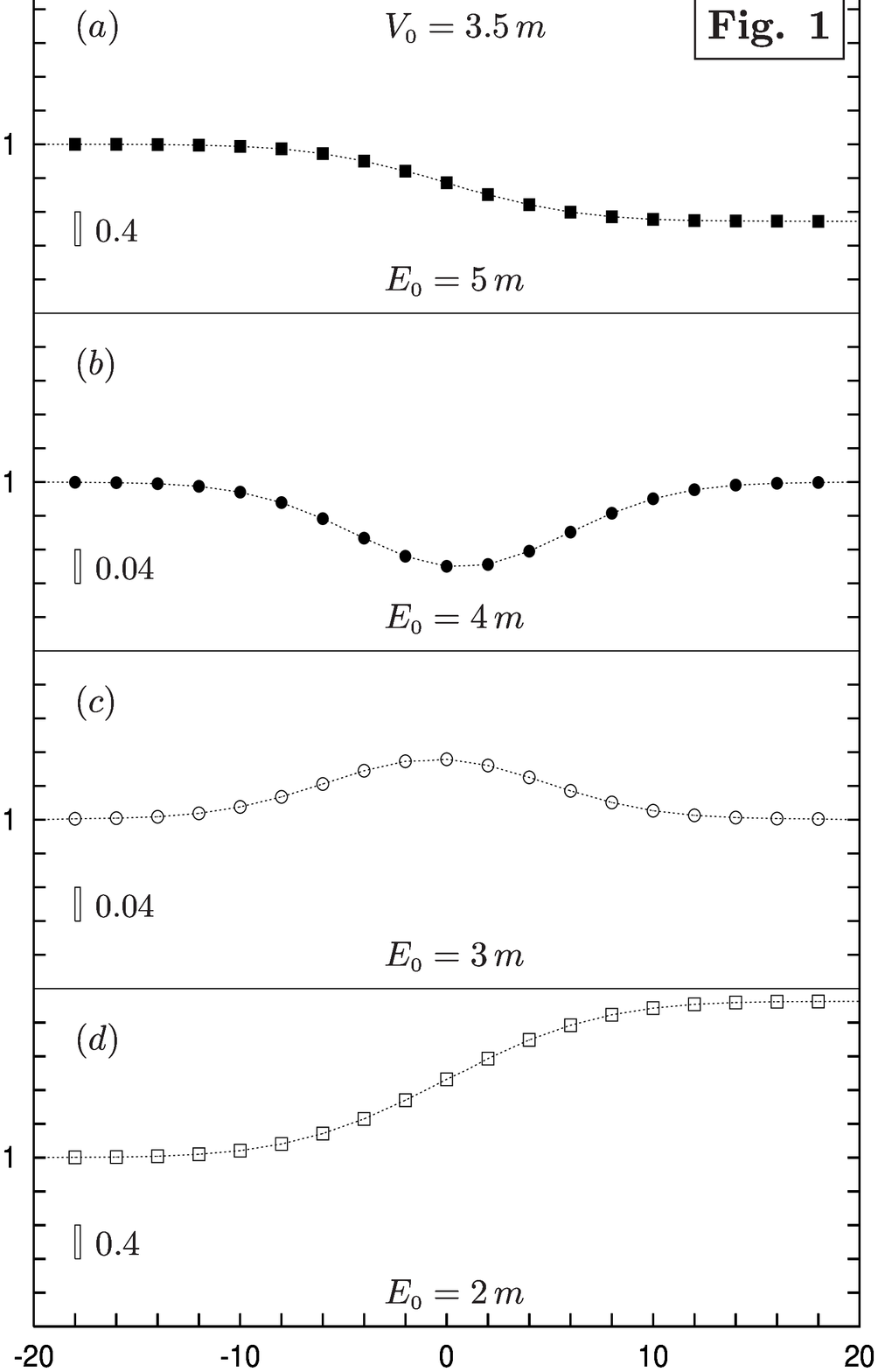}
\vspace*{-1.8cm}
 \caption{The ratio of particles in region I to those of an incident gaussian wave
 packet peaked at $E_{\0}$ and with localization $md=10$ as a function of
 time. The value of $V_{\0}$ has been set to $3.5\,m$. The four
 energy zones correspond to (a) diffusion [$E_{\0}>V_{\0}+m$], (b) Dirac tunneling
[$V_{\0}<E_{\0}<V_{\0}+m$], (c) Klein tunneling
[$V_{\0}-m<E_{\0}<V_{\0}$], and (d) Klein [$E_{\0}<V_{\0}-m$].}
\end{figure}

\newpage

\begin{figure}[hbp]
\hspace*{-2.5cm}
\includegraphics[width=19cm, height=24cm, angle=0]{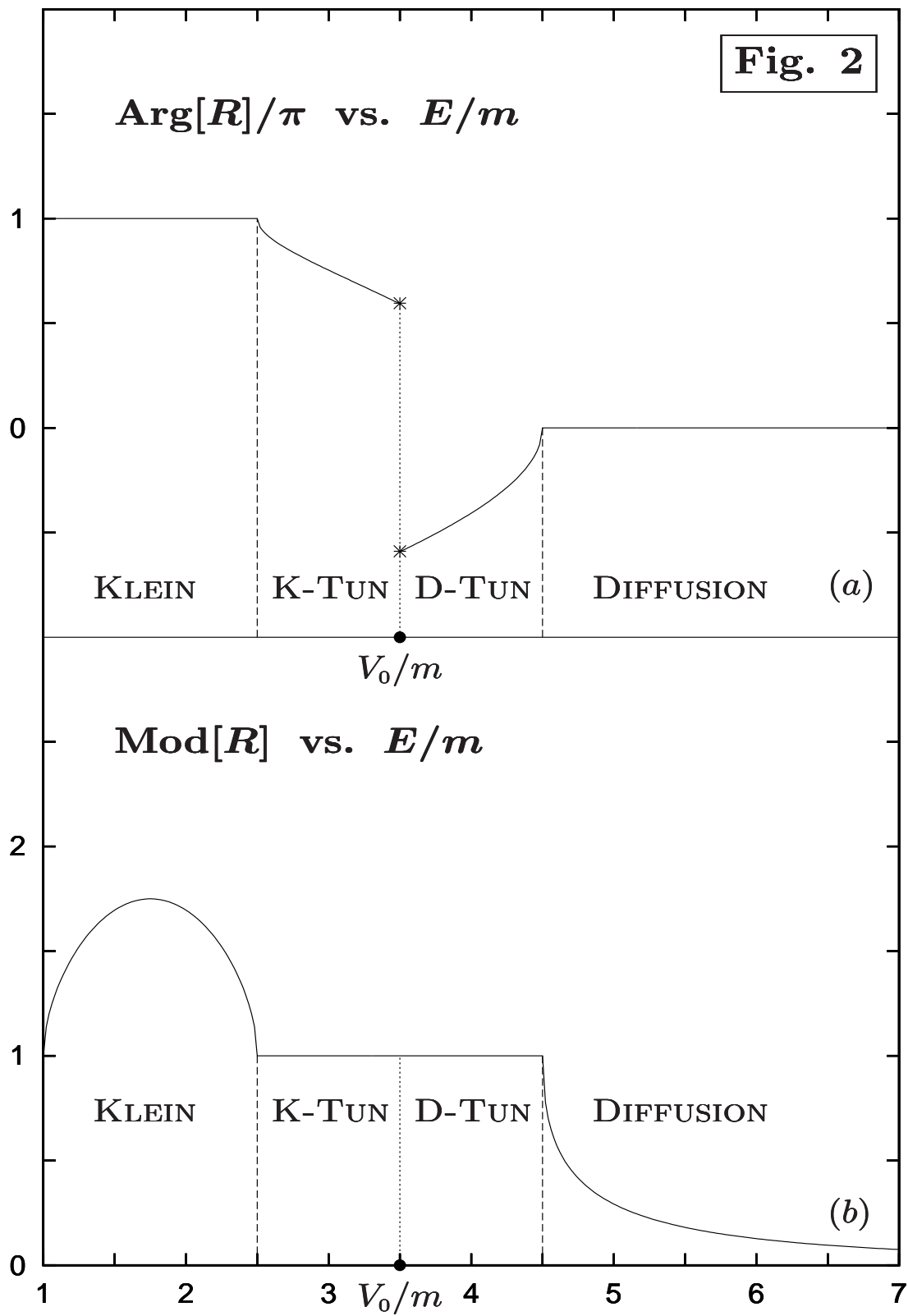}
\vspace*{-1.8cm}
 \caption{The phase and modulus of $R$ as a function of $E/m$ for $V_{\0}=3.5\,m$.
 The discontinuity in the phase is at $E_{\0}=V_{\0}$ and the peak value of the modulus
 in the Klein zone is at $E_{\0}=V_{\0}/2$.}
\end{figure}

\end{document}